\documentclass[twocolumn,showpacs,amsmath,amssymb,prl,nofootinbib]{revtex4-1}

\usepackage[pdftex]{graphicx} 
\usepackage{natbib}

\def\bea{\begin{eqnarray}}
\def\eea{\end{eqnarray}}
\def\ben{\begin{equation}}
\def\een{\end{equation}}
\def\benu{\begin{enumerate}}
\def\enu{\end{enumerate}}

\def\sss{\scriptscriptstyle\rm}

\def\xc{_{\sss XC}}

\def\ee{_{\rm ee}}

\begin{document}

\title{One-dimensional Continuum Electronic Structure with the Density Matrix Renormalization Group and Its Implications for Density Functional Theory}
\author{E.M. Stoudenmire}
\affiliation{Department of Physics and Astronomy, University of California, Irvine, CA 92697}
\author{Lucas O.\ Wagner}
\affiliation{Department of Physics and Astronomy, University of California, Irvine, CA 92697}
\author{Steven R.\ White}
\affiliation{Department of Physics and Astronomy, University of California, Irvine, CA 92697}
\author{Kieron Burke}
\altaffiliation[Also at ]{Department of Chemistry, University of California, Irvine, CA 92697}
\affiliation{Department of Physics and Astronomy, University of California, Irvine, CA 92697}
\date{\today}

\begin{abstract}
We extend the density matrix renormalization group to compute 
exact ground states of continuum many-electron systems in one dimension with long-range interactions. 
We find the exact ground state of a chain
of 100 strongly correlated artificial hydrogen atoms. 
The method can be used
to simulate 1d cold atom systems and to study density functional theory in an exact setting.
To illustrate, we find an interacting, extended system which is 
an insulator but whose Kohn-Sham system is metallic.
\end{abstract}


\maketitle

For electronic structure calculations, these are the best of times and the worst of times.
When correlations are weak, density functional theory (DFT) makes it possible to 
tackle extremely realistic Hamiltonians and large system sizes with reasonable 
accuracy \cite{Martin:2004,*Fiolhais:2003}.
For strongly correlated systems, there exist powerful and controllable numerical methods
\cite{Corboz:2011,*Lauchli:2011,*Sandvik:2010} for simulating lattice Hamiltonians,
such as the Hubbard model.
However, few numerical tools can treat the combination of
strongly correlated electronic systems and realistic microscopic Hamiltonians. 
In the strongly correlated regime, DFT approximations are neither systematic
nor controllable, often leading to unrestrained parameter multiplication and
empiricism.  Model Hamiltonians rely on the arbitrary truncation of terms that
may be crucial in tipping the balance between competing phases.  Attempts to
bridge the gap between realistic Hamiltonians and strong correlation
techniques, such as dynamical mean field theory coupled to DFT \cite{Anisimov:1997,Lichtenstein:1998}, 
may contain both arbitrary truncations and a less than
ideal treatment of correlations.

Therefore we would like to study DFT in an exact setting to see
how density functional approximations break down and 
whether new approximations contain the right physics.
But very few continuum, three-dimensional, long-range 
interacting systems can be easily treated exactly.
Here, we show that by studying one dimensional (1d) systems instead, we can
treat realistic Hamiltonians and strong electron correlations essentially exactly, 
even for a very large number of atoms. 
Because they preserve the continuum, our 1d models mimic key features of three-dimensional
reality surprisingly well \cite{Wagner:2012}.

\begin{figure}[t]
\includegraphics[width=\columnwidth]{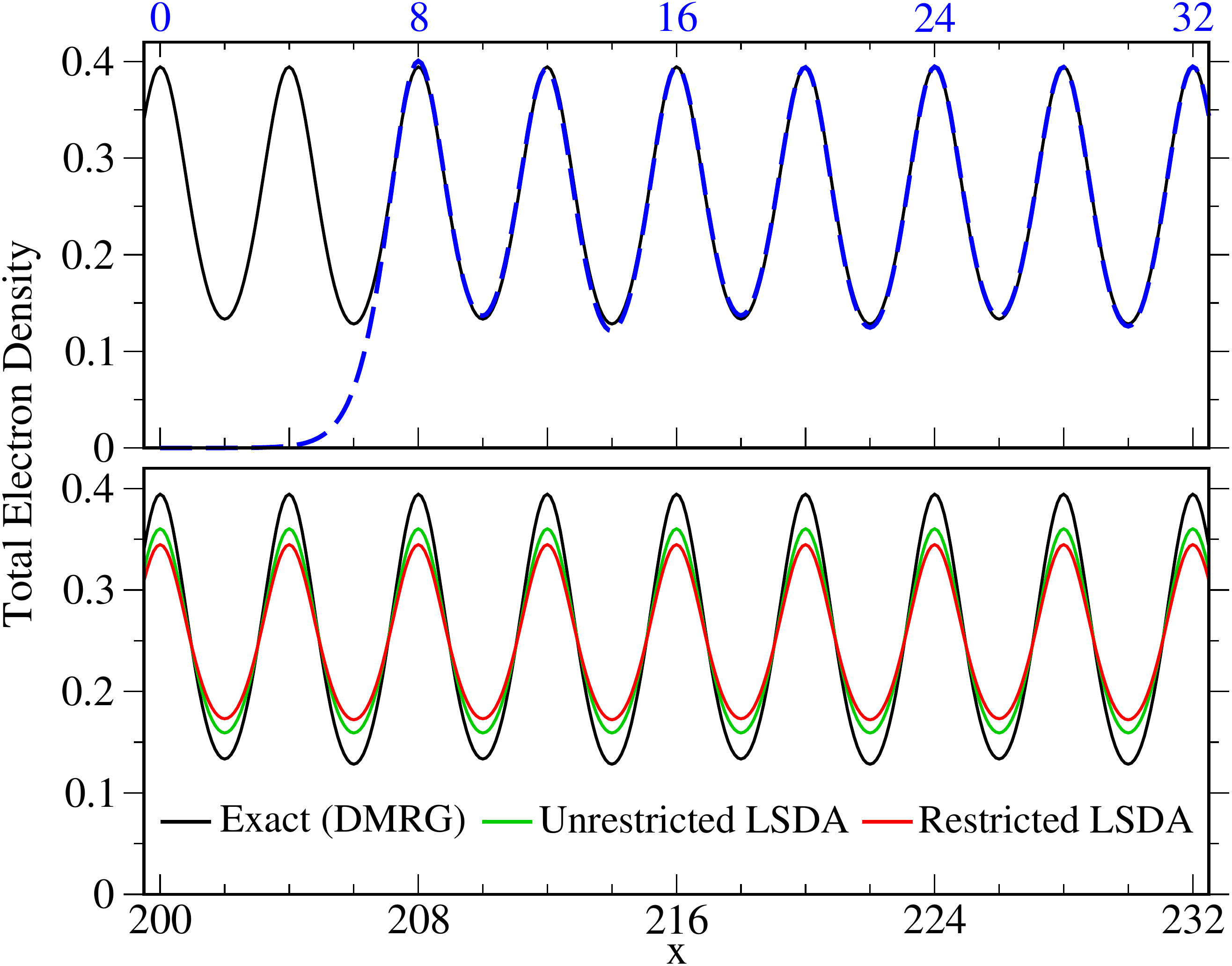}
\caption{The exact ground state density of a chain of 100 widely separated (strongly correlated) artificial atoms.
The total length of the system is $L=420$ in atomic units (4200 grid sites with a spacing of 0.1).
The upper panel shows the electron density of a central region superimposed with the density at the left edge 
(the dashed blue curve with corresponding $x$ above). 
The lower panel compares the exact electron density to DFT predictions within the local spin density approximation.} 

\label{fig:100atom}
\vskip -0.5cm
\end{figure}

Our approach is based on the density matrix renormalization group (DMRG) \cite{White:1992,*White:1993a},
the most powerful of the strongly correlated techniques for 1d lattice models.
Here we extend DMRG to treat continuum electron systems with long-range interactions. 
This new approach retains DMRG's exponential convergence and near linear scaling with system size. 
As an example, we present a near exact calculation of a
system with 100 strongly interacting pseudo-hydrogen atoms (Fig.~\ref{fig:100atom}).

A key motivation for this method is to study DFT in an exact 
setting, both when correlations are strong and near the thermodynamic limit.
Generically, 1d systems have strong quantum fluctuations, making them 
an especially rigorous test of DFT approximations; they can also be pushed
to large size with less effort.
As in Fig.~\ref{fig:100atom}, we can easily compare various DFT approximations with exact
results for extended systems.
We can also compute exact quantities appearing in the DFT formalism; 
for example, we show below that a gapped interacting system can 
nevertheless have a Kohn-Sham gap which is exactly zero (a Mott insulator \cite{Mott:1937,*Mott:1949}). 
DMRG also offers new ways to characterize electronic structure models
using quantum information concepts, such as the bipartite entanglement entropy.
Finally, 1d continuum Hamiltonians can be realized exactly in cold atom systems \cite{Liang:2011,Cazalilla:2011b}.

Presently, real-space DMRG methods for solid-state applications are designed only to work with lattice models.
Each site of such a model can be thought of as a Wannier function centered on an atom. One way of generalizing
this picture to make the Hamiltonian more realistic is to expand in a set of basis functions;
DMRG has become a powerful technique for the quantum chemistry of small molecules based on this idea
\cite{White:1999,Chan:2011}. Here we proceed in a more flexible direction which 
does not depend on a choice of basis and has optimal scaling of calculation time with system
size: we represent the continuum in 1d with a real space grid. 
The continuum Hamiltonians of interest can be written as
\begin{align}
H  & =  \sum_\sigma \int_x \psi^\dagger_\sigma(x) \left[ -\frac{1}{2} \frac{\partial^2}{\partial x^2} - \mu \right] \psi_\sigma(x) \nonumber \\
& \mbox{} + \int_x v(x)\, n(x) + \frac{1}{2} \int_{x,x^\prime} v\ee(x-x^\prime)\, n(x)\, n(x^\prime) \ . \label{eqn:continuum_H}
\end{align}
where $\psi^\dagger_\sigma(x)$ creates an electron of spin $\sigma$ at position $x$; 
$v$ and $v\ee$ are local and electron-electron potentials, respectively; and
$n$ is the electron density operator.
We introduce a grid spacing $a$, obtaining a discretized Hamiltonian
\begin{align}
H & = \sum_{j,\sigma}  \frac{-1\ \ }{2 a^2} (c^\dagger_{j\sigma} c_{j+1\sigma} + c^\dagger_{j+1\sigma} c_{j\sigma}) -\tilde\mu\, n_{j\sigma} \nonumber \\
& \mbox{} + \sum_j v^j \, n_j + \frac{1}{2} \sum_{i,j} v\ee^{ij}\ n_i \, (n_j-\delta_{ij})  \label{eqn:discrete_H}
\end{align}
where \mbox{$\tilde\mu = \mu-1/a^2$}, \mbox{$v^j = v(j\, a)$} and \mbox{$v\ee^{ij} = v\ee(|i-j|\, a)$}.
The $\delta_{ij}$ in the last term prevents an unphysical self interaction.
For technical reasons we work with open boundaries rather than periodic, and extend the grid well past the
edge atoms.
Finite grid spacing errors can be reduced arbitrarily by reducing $a$; convergence can be  accelerated 
by using a higher order discretized derivative \cite{Wagner:2012}. Here, we fix $a=0.1$ in atomic units.

A  different approach was proposed recently by Verstraete and Cirac \cite{Verstraete:2010},
who showed how to define and optimize matrix product states directly in the continuum 
limit in the context of quantum field theories. 

Working {\it efficiently}  with such a Hamiltonian represents an unusual challenge for DMRG.
Normally, one is most concerned with the number of states per block $m$ needed to represent
the ground state.  The number of sweeps $N_S$ needed to converge to the ground state
is usually quite small ($\sim\! 2\!-\!5$) for 1d systems. Here, the reverse can happen: 
a small grid \mbox{spacing `$a$'} relative to the interatomic separation 
can lead to energy scales that differ by orders of magnitude,
greatly increasing $N_S$ but not affecting $m$ significantly. 
Fortunately, while convergence with $m$ reflects the entanglement of the system, an
inherent property, many approaches can be tried to reduce $N_S$. 
We have found a particularly efficient acceleration approach based on a real-space RG procedure
which produces a supplementary grid with a much coarser spacing and lower energy scales such that
$N_S$ can be made small; after sweeping on this grid we map the wave function back onto
the fine grid for further sweeps. 
This procedure, which we hope to discuss in a future publication,
introduces no additional approximations; it merely reduces the computational time
to reach a converged ground state.

Besides the wide range of energy scales, another challenge for DMRG is 
the presence of long-ranged interactions. The simplest approach, dealing with $N_g^2$ terms
in a sweep over $N_g$ grid points, would scale as $N_g^3$.  A more efficient approach using intermediate
operators, as used with DMRG for quantum chemistry in a basis set, would scale as $N_g^2$.
We utilize a much more efficient approach than either of these by using
a representation of the Hamiltonian as a matrix product operator (MPO).
Finite bond dimension MPOs naturally encode exponentially decaying interactions \cite{McCulloch:2007}.
Interactions with a power law decay can be approximated to high accuracy by fitting
to a sum of $N_{\rm MPO}$ exponentials (usually $N_{\rm MPO} \leq 50$ is 
sufficient to obtain an accuracy of $10^{-5}$) \cite{Pirvu:2010}. 
Thus the DMRG calculation time is nominally linear in $N_g$; specifically, 
it is proportional to $N_S N_g N_{\rm MPO}\, m^3$.  Deviations from a purely linear computational effort with
the overall system size can come from a dependence of $N_S$ or $m$ with system size; in practice
we find that the RG acceleration procedure keeps $N_S$ small. The behavior of $m$ is well understood
for short-ranged model Hamiltonians: for noncritical systems, $m$ is independent of system size.
For critical systems, with power-law decaying correlations, $m$ grows only logarithmically with length. 
Hence, we expect only a slightly worse-than-linear computational effort with system size 
in the worst case.

As a simple setting for exploring strong many-body correlation effects, we consider
chains of one-dimensional ``soft hydrogen" atoms.
Each atom consists of a single electron in a soft-Coulomb potential well of the form
\ben
v_{\text{atom}}(x) = -1/{\sqrt{x^2 + 1}} \:, \label{eqn:softCoulomb}
\een
and we use atomic units throughout.
We also include repulsive interactions defined by 
\mbox{$v\ee(x-x') = -v_{\text{atom}}(|x-x'|)$}. 
Using bare Coulomb interactions would lead to an ill-defined model in 1d;
this potential is a standard choice for avoiding such complications and 
has been used to study molecules in 
intense laser fields \cite{Eberly:1989,Thiele:2008}.
We are also fortunate to benefit from the work of Ref.~\onlinecite{Helbig:2011},
which provides a parameterization of the 1d local spin density approximation (LSDA) for just such an interaction.
We note that by using state-of-the-art DMRG it is also possible to simulate 
a chain of real hydrogen atoms \cite{Hachmann:2006}. But our purely 1d setting allows us to study
a wider variety of systems and, in the future, explore dynamical and finite-temperature effects.

\begin{figure}[b]
\includegraphics[width=\columnwidth]{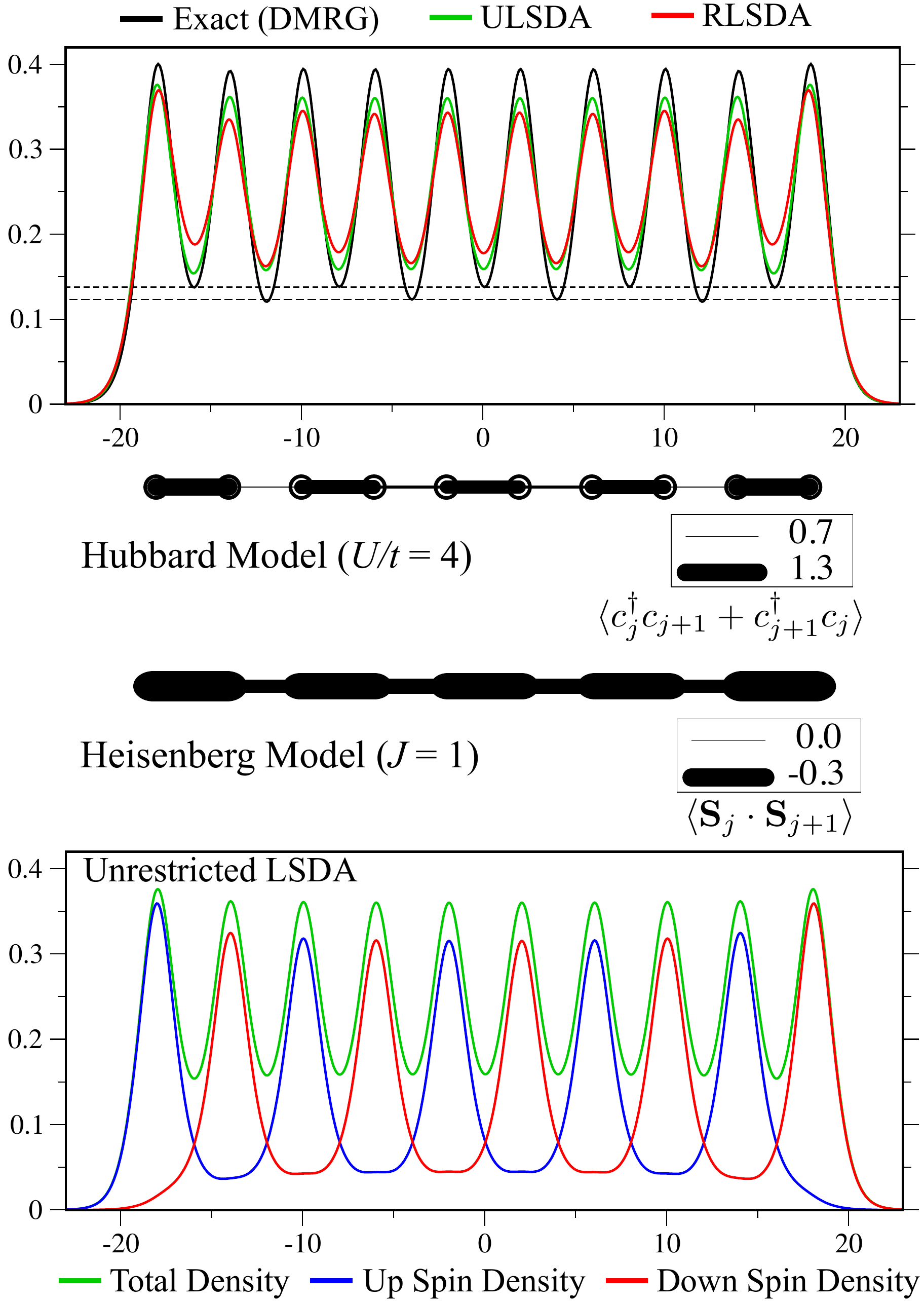}
\caption{Spontaneous dimerization of the density for a chain of 10 soft hydrogen
atoms with interatomic spacing $b=4$ (dashed lines are a guide to the eye). The upper panel compares the densities predicted by DFT within 
the LSDA; the lower panel shows the spin densities for unrestricted LSDA. Also shown
is the expectation value of the kinetic energy $\langle c^\dagger_j c_{j+1} + c^\dagger_{j+1} c_j\rangle$ for a Hubbard model with $U/t=4$ 
and the exchange energy $\langle \mathbf{S}_j \cdot \mathbf{S}_{j+1}\rangle$ for the Heisenberg model on 10 lattice sites. 
The thickness of the lines indicates the magnitude of these quantities on each bond.}
\label{fig:dimers}
\vskip -0.5cm
\end{figure}

To demonstrate of the power of our approach, we display in
Fig.~\ref{fig:100atom} the exact ground state density of a chain of one hundred
artificial atoms with long-range interactions, which took a few days of computer time on a single
workstation. 
Representing the ground state accurately required keeping about $m=200$ states.
The relative energy error in the many-body solution from DMRG is of order $10^{-6}$.
The relative errors due to the finite grid spacing and finite number of exponentials $N_{\rm MPO}$ used to
fit $v\ee$ are larger, of order $10^{-4}$, but are well understood and easily reduced if necessary.
Also shown in Fig.~\ref{fig:100atom} are DFT calculations within both the restricted and unrestricted LSDA \cite{Helbig:2011}.
For this system, both DFT approaches make substantial errors. 
In terms of lattice models, one would represent this system with either a half-filled 
Hubbard chain or an antiferromagnetic Heisenberg chain.  
Both models are critical with power law decaying spin-spin correlations; 
a noncritical system would have been easier for DMRG. 
It is not surprising that the local DFT approximations cannot capture
the quasi-long-range spin correlations, with the unrestricted LSDA predicting long-range
antiferromagnetism (similar to Fig.~\ref{fig:dimers}).
It is somewhat surprising that even the total  density from LSDA deviates
strongly from the exact DMRG results.  The total energy from each LSDA
calculation is off by about 1\%.

In Fig.~\ref{fig:dimers}  we show a system which reveals weaknesses of both 
approximate DFT and of model Hamiltonian approaches. The figure shows the exact 
ground state density of ten atoms 
with interatomic spacing $b=4$. 
The edges induce a staggered pattern of strong and weak bonds which decays slowly into the bulk, 
and is therefore significant throughout this small system.
We can understand the staggered behavior from a 10 site Hubbard model (at
half filling with $U/t=4$ chosen arbitrarily) 
or a 10 site Heisenberg model. The Heisenberg ground state has resonating valence bond
character; in a perfect near-neighbor RVB state, the edges would suppress all resonance
and drive the weak bonds to zero. The actual Heisenberg ground state has longer range
resonances which reduce these effects. In the Hubbard model, the strong exchange
bonds show up as bonds with lower kinetic energy.  However, neither lattice model
reveals the increased electron density on the strong bonds, stemming from the strong hopping. 
These models might be improved by bond-dependent interactions $t$ and $J$.
The LSDA calculations capture even fewer properties of the true ground state.
Unrestricted LSD predicts an energy
$-11.364$ which is close to the exact energy $-11.496$, but
no staggered bond density and breaks spin symmetry, producing
a long-ranged antiferromagnetic state as shown in the lower panel of Fig.~\ref{fig:dimers}.
Restricted LSDA captures the staggered density pattern qualitatively, but gives
a slightly higher energy $-11.323$ and fails to reproduce the correct
local spin correlations since its wave function is a Slater determinant of
extended orbitals.  
The artificial symmetry breaking of LSDA can be understood as a frozen spin fluctuation \cite{Perdew:1995},
but the exact functional yields a singlet ground state.

\begin{figure}[t]
\includegraphics[width=\columnwidth]{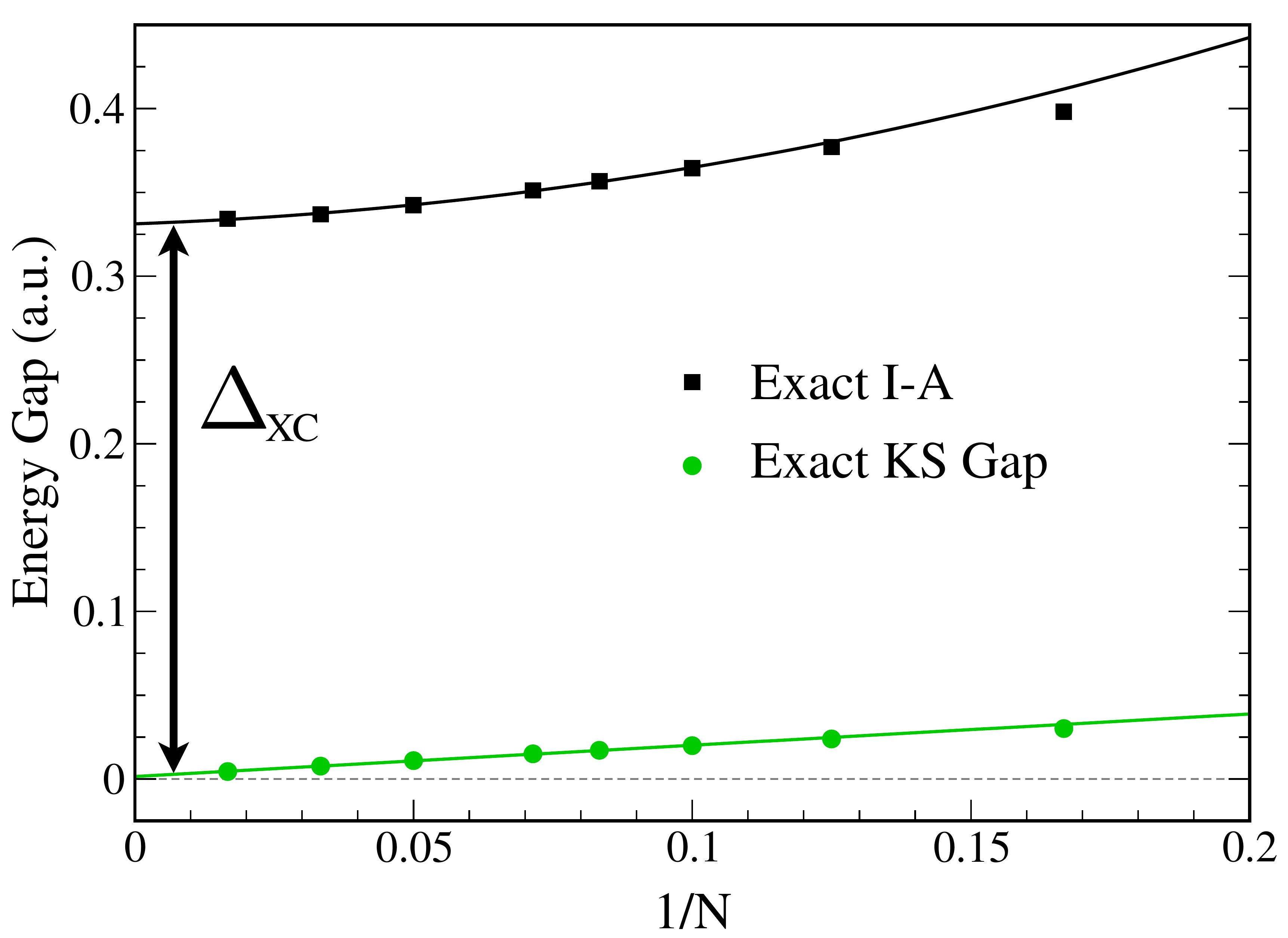}
\caption{Exact gaps for chains of $N$ soft hydrogen atoms with atomic 
    separation $b=4$ (error bars are less than symbol sizes). The upper curve is a quadratic fit of 
    exact gaps of the largest six systems and extrapolates to a finite value \mbox{$E_g \simeq 0.33$}.
    The exact Kohn-Sham gaps, in contrast, extrapolate to zero showing that for $N\rightarrow\infty$
    the true KS system is metallic (lower curve is a linear fit of exact KS gaps of the largest six systems).}
\label{fig:gaps}
\end{figure}

Not only can we compare our exact results to DFT approximations, but we can use them to
investigate fundamental questions about DFT itself.
The fundamental (charge) gap is $E_g=(I-A)$
where $I$ is the ionization potential and $A$ the electron affinity.
In Fig.~\ref{fig:gaps}, we compute $E_g$ for chains of soft hydrogen atoms 
with spacing $b=4$ for large systems up to $N=60$ atoms ($\sim 2500$ grid sites).
Extrapolation shows the $N\rightarrow\infty$ system to be an insulator.
We also compute the exact Kohn-Sham (KS) gap for each $N$ by inverting the density of the neutral
system to obtain the KS potential and its single particle energies.
(Given an interacting system, the KS system is the unique non-interacting system with the same
density \cite{Kohn:1965}.)
In the thermodynamic limit, the KS gaps extrapolate to zero, so that the exact $N\rightarrow\infty$ KS system is a metal.
This is consistent with the fact that each finite KS system in Fig.~\ref{fig:gaps} has one electron per unit cell and thus a half-filled
band (in contrast to the unrestricted LSDA which breaks spin symmetry for this system). 

The discrepancy between the KS and exact gap was long ago identified \cite{Perdew:1982} with the exchange-correlation derivative discontinuity in DFT:
$E_g = \Delta_s + \Delta\xc$ where $\Delta_s$ is the
KS gap, that is, the HOMO-LUMO energy difference for the neutral KS system.
Approximate functionals such as LSDA that are continuous in particle number miss this effect entirely.
The LSDA KS gaps are almost identical to the exact ones shown in Fig.~\ref{fig:gaps}, but the LSDA fundamental gap drops from 
close to $E_g$ for small $N$ to near zero at large $N$ (details reported elsewhere).

Previous calculations have found $\Delta\xc$ for semiconductors
\cite{Knorr:1992,*Knorr:1994,Gruning:2006d} with finite KS gaps $\Delta_s$, but
our system's gap is entirely due to $\Delta\xc$, underscoring its
importance for strong correlation physics.
Our results rely on no uncontrolled approximations and so demonstrate unambiguously the
behavior of Mott insulators in DFT.  
Present DFT research on this issue focuses on 
extracting accurate $E_g$ from semilocal functional calculations \cite{Chan:2010,Zheng:2011}.

\begin{figure}[t]
\includegraphics[width=\columnwidth]{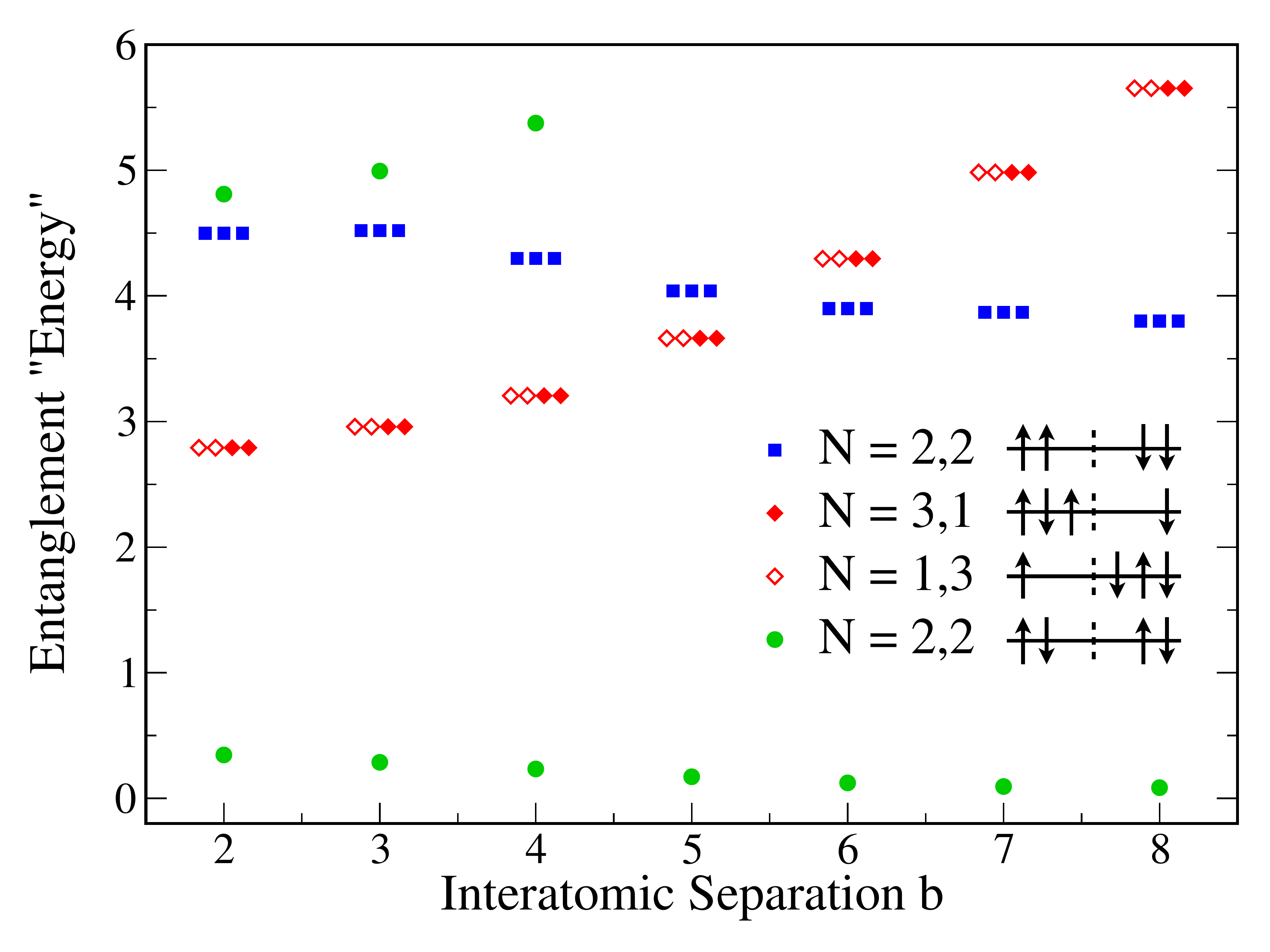}
\caption{Entanglement spectrum at the center of interacting 4-atom chains with various interatomic separations $b$.
\mbox{$N = (N_L, N_R)$} refers to the number of electrons to the left and right of the cut for each
density matrix eigenstate. 
The states with $N_L=3,1$ primarily correspond to charge fluctuations while those with $N_L=2$
to spin fluctuations.}
\label{fig:4atom_ent_spectrum}
\end{figure}

The onset of strong correlation with increasing bond length is often identified with the 
Coulson-Fischer point \cite{Coulson:1949}, where
an unrestricted Hartree-Fock calculation spontaneously breaks spin symmetry.
A different way to distinguish strong from weak correlation is through the entanglement spectrum,  
readily accessible in DMRG.
Defining the left reduced density matrix 
\mbox{$\rho_L = \text{Tr}_R |\Psi\rangle \langle \Psi |$}, where the trace is
over all grid sites in the right half of the system, the entanglement spectrum
consists of the energies of the entanglement Hamiltonian 
\mbox{$H_E = -\ln \rho_L$} \cite{Turner:2011}.
The most probable density matrix eigenstates are those in the low ``energy" part of the spectrum.
By classifying these states according to their particle number $N_L$, we can understand the dominant
quantum fluctuations of the ground state.
Figure~\ref{fig:4atom_ent_spectrum} shows the entanglement spectrum 
at the center of a series of four-atom chains with increasing interatomic separation.
A sharp crossover at $b\simeq 5.5$, where the probability for charge fluctuations 
drops below that of pure spin fluctuations, signals the onset of strongly correlated behavior.

Many oxide materials of current interest are too strongly correlated for present DFT methods, but crucial properties 
must be calculated to an accuracy far beyond that of simple model Hamiltonians. The method described here provides 
a new, alternative route to studying strongly correlated systems. All existing approximations, from heuristic corrections to standard 
functionals, such as LDA+U \cite{Anisimov:1991}, to methods developed for lattice models, such as dynamical mean 
field theory \cite{Georges:1996}, can be applied and tested more easily, thoroughly, and accurately in the present setting.  
Because our 1d world captures a feature crucial to density functional approximations, namely the continuum instead of a lattice, such 
studies should provide the insight needed to construct more accurate density functionals for real strongly-correlated materials.

\emph{Note added:} After completing this Letter, we became aware of Ref.~\onlinecite{Dolfi:2012} which is 
similar in spirit to our real-space RG method for accelerating continuum DMRG.

\acknowledgements{We gratefully acknowledge DOE grant DE-FG02-08ER46496 (KB, LW and SW)  and
NSF grant DMR-0907500 (ES and SW) for supporting this work.}

\bibliography{dmrg_dft}

\end{document}